\documentclass[
twocolumn,
aps,nofootinbib,showpacs,showkeys,preprint
tightenlines,preprintnumbers,
,superscriptaddress
] {revtex4}

\usepackage{epsf,epsfig,subfigure,graphicx,amsmath,amssymb}
\usepackage{color}
\newcommand{\dis}[1]{\begin{equation}\begin{split}#1\end{split}\end{equation}}

\def\lsim{\lower.7ex\hbox{$\;\stackrel{\textstyle<}{\sim}\;$}}

\def\ie{{\it i.e.}\ }

\begin{document}
%\draft
%\preprint{SNUTP 11-008}
\title{A simple expression of the Jarlskog determinant}
\author{Jihn E.  Kim}
\address{GIST College, Gwangju Institute of Science and Technology, Gwangju 500-712, Korea
}
\author{Min-Seok Seo
}
\address{Department of Physics and Astronomy and Center for Theoretical
 Physics, Seoul National University, Seoul 151-747, Korea\\
}

\begin{abstract}
Making the whole determinant of the CKM matrix $V$ real, the imaginary part of any one term of the determinant of $V$ (e. g. $|{\rm Im}\,V_{31}V_{22}V_{13}|$) is the Jarlskog determinant $J$. We also point out that the maximality of the weak CP violation is a physical statement.
\end{abstract}


 \maketitle

%%%%%%%%%%%%%%%%%%%%%%%%%%%%%%%%%%%%%%%%%%%%%%%%%%%%%%%%%%%%%%%%%%%
%%%%%%%%%%%%%%%%%%%%%%%%%%%%%%%%%%%%%%%%%%%%%%%%%%%%%%%%%%%%%%%%%%%
The Jarlskog determinant is \cite{Jarlskog85}
\dis{
J = \frac{-{\rm Det.}\, C}{2F(m_{t,c,u})F(m_{b,s,d})}
}
where $F(m_{t,c,u})=(m_t-m_c)(m_t-m_u)(m_c-m_u)$ and $F(m_{b, s,d})=(m_b-m_s)(m_b-m_d)(m_s-m_d)$,
and $C$ is
\dis{iC=[M_u, M_d],
}
where $M_{u,d}=L^{(u,d) \dagger}M^{0(u,d)} L^{(u,d)}$ with the diagonal $M^{0(u,d)}$. It can be also written as $iC=[M_u M_u^\dagger, M_d M_d^\dagger].$
Then, one can show that  \cite{JarlskogBook},
\dis{
{\rm Det.\,} C=-2J F(m_{t,c,u}^2)F(m_{b,s,d}^2).
}
This Jarlskog determinant is product of two elements of $V$ and two elements of $V^*$, just twice the area of the Jarlskog triangle.

If the determinant is real, \ie for the unit determinant it can be written as
 \dis{ 1=&
 V_{11}V_{22}V_{33}-V_{11}V_{23}V_{32}+V_{12}V_{23}V_{31}\\
& -V_{12}V_{21}V_{33}+V_{13}V_{21}V_{32}-V_{13}V_{22}V_{31}.\label{eq:detdetail}
}
Multiplying $V_{13}^*V_{22}^*V_{31}^*$ on both sides, we obain
\dis{V_{13}^* &V_{22}^*V_{31}^* =|V_{22}|^2V_{11}V_{33}V_{13}^*V_{31}^*-V_{11}
 V_{23}V_{32}V_{13}^*V_{31}^*V_{22}^*\\
 &+|V_{31}|^2V_{12}V_{23}V_{13}^*V_{22}^* -V_{12}V_{21}V_{33}V_{13}^*V_{31}^*V_{22}^*\\
 &+|V_{13}|^2V_{21}V_{32}V_{31}^*V_{22}^*-|V_{13}V_{22}V_{31}|^2.\label{eq:detmultV*}
 }
We will show that the imaginary part of the left-hand side (LHS) of Eq. (\ref{eq:detmultV*}), \ie  $|{\rm Im}\,V_{31}V_{22}V_{13}|$, is the Jarlskog determinant $J$. Firstly, consider the second term on the right-hand side (RHS), $-V_{11}V_{23}V_{32} V_{13}^*V_{31}^*V_{22}^*$. It contains a factor $V_{32}V_{22}^*$, which is equal to $-V_{31}V_{21}^*-V_{33}V_{23}^*$ by the unitarity of $V$. Then, $-V_{11}V_{23}V_{32}V_{13}^*V_{31}^*V_{22}^*= V_{11}V_{23} V_{13}^*V_{21}^*|V_{31}|^2 +V_{11}V_{33}V_{13}^*V_{31}^*|V_{23}|^2.$ Especially, the second term $V_{11}V_{33}V_{13}^*V_{31}^*|V_{23}|^2$ combines with the first term of Eq. (\ref{eq:detmultV*}), $|V_{22}^2|V_{11}V_{33}V_{13}^*V_{31}^*$, to make $(1-|V_{21}|^2)V_{11}V_{33}V_{13}^*V_{31}^*$. Second, note that the fourth term  on the RHS of Eq. (\ref{eq:detmultV*}), $-V_{12}V_{21}V_{33}V_{13}^*V_{31}^*V_{22}^*$ containing the factor $V_{33}V_{31}^*=-V_{23}V_{21}^*-V_{13}V_{11}^*$, can be rewritten as $-V_{12}V_{21}V_{33}V_{13}^*V_{31}^*V_{22}^*=V_{12}V_{23}V_{13}^*V_{22}^*|V_{21}|^2
+V_{12}V_{21}V_{11}^*V_{22}^*|V_{13}|^2.$
Here, the first term $V_{12}V_{23}V_{13}^*V_{22}^*|V_{21}|^2$ combines with the third term on the RHS of Eq. (\ref{eq:detmultV*}), $|V_{31}|^2V_{12}V_{23}V_{13}^*V_{22}^*$, to make $(1-|V_{11}|^2)V_{12}V_{23}V_{13}^*V_{22}^*$.

In sum, Eq. (\ref{eq:detmultV*}) can be rewritten as
 \dis{
  &V_{13}^*V_{22}^*V_{31}^* =(1-|V_{21}|^2)V_{11}V_{33}V_{13}^*V_{31}^* \\
 &~+V_{11}V_{23} V_{13}^*V_{21}^*|V_{31}|^2 +(1-|V_{11}|^2)V_{12}V_{23}V_{13}^*V_{22}^*\\
 &~+|V_{13}|^2(V_{12}V_{21}V_{11}^*V_{22}^*+V_{21}V_{32}V_{31}^*V_{22}^*)\\
 &~-|V_{13}V_{22}V_{31}|^2.
 \label{eq:detmult3}
 }
Now, the unitarity plays an important role in simplifying this expression. Let the imaginary part of $V_{11}V_{33}V_{13}^*V_{31}^*$ be $J$. From $V_{11}^*V_{13}+V_{21}^*V_{23}+V_{31}^*V_{33}=0$, we know $|V_{11}|^2|V_{13}|^2+V_{11}V_{23}V_{13}^*V_{21}^*+V_{11}V_{33}V_{13}^*V_{31}^*=0$; so the imaginary part of $V_{11}V_{23}V_{13}^*V_{21}^*$ is $-J$. From $V_{11}V_{31}^*+V_{12}V_{32}^*+V_{13}V_{33}^*=0$, we have $V_{11}V_{33}V_{13}^*V_{31}^*+V_{12}V_{33}V_{32}^*V_{13}^*+|V_{13}^*V_{33}|^2=0$.  And, from $V_{12}^*V_{13}+V_{22}^*V_{23}+V_{32}^*V_{33}=0$,  we have $V_{12}V_{33}V_{32}^*V_{13}^*+V_{12}V_{23}V_{22}^*V_{13}^*+|V_{12}^*V_{13}|^2=0$. These two combine to show that the imaginary part of $V_{12}V_{23}V_{22}^*V_{13}^*$ is $J$. On the other hand, from $V_{11}^*V_{12}+V_{21}^*V_{22}+V_{31}^*V_{32}=0$, we know $V_{21}V_{32}V_{22}^*V_{31}^*+V_{12}V_{21}V_{11}^*V_{22}^*+|V_{21}^*V_{22}|=0$; so
the imaginary part of $(V_{21}V_{32}V_{22}^*V_{31}^* +V_{12}V_{21}V_{11}^*V_{22}^*)$ is zero, leading to the conclusion that the imaginary part of the RHS of Eq. (\ref{eq:detmult3}) is
$[(1-|V_{21}|^2)-|V_{31}|^2+(1-|V_{11}|^2)]J=J$.
Therefore, the imaginary part of  $V_{13}^*V_{22}^*V_{31}^*$ (the LHS of Eq. (\ref{eq:detmult3})) is $J$.  It is {\it the imaginary part of any one element among the six components of determinant of $V$} as explicitly shown in \cite{KimSeo11} when the whole determinant is made real as defined in Eq. (\ref{eq:detdetail}). This simplifies how the weak CP violation is scrutinized just from looking at the CKM matrix elements. We remind that the reality of the determinant of the quark mass matrix, which is related to the reality condition of $V$, is important in the strong CP study \cite{KimRMP10}.

%%%%%%%%%%%%%%%%%%%%%%%%%%%%%%%%%%%%%%%%%%%%%%%%%%%%%%%%%%%%%%%%%%%%%%%%%%%%%%%%%%%%
\begin{figure}[!t]
  \begin{center}
  \begin{tabular}{c}
   \includegraphics[width=0.35\textwidth]{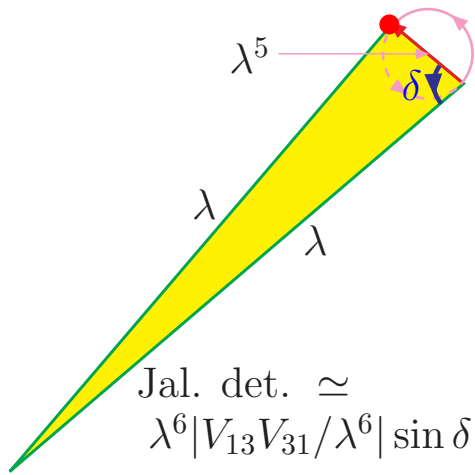}
   \end{tabular}
  \end{center}
 \caption{The Jarlskog triangle. This triangle is for two long sides of $O(\lambda)$. Rotating the $O(\lambda^5)$ side (the red arrow), the CP phase $\delta$ changes.
  }
\label{fig:JTriangle}
\end{figure}
%%%%%%%%%%%%%%%%%%%%%%%%%%%%%%%%%%%%%%%%%%%%%%%%%%%%%%%%%%%%%%%%%%%%%%%%%%%%%%%%%%%%%

We can argue that the maximality of the weak CP violation is a physical statement. The physical magnitude of the weak CP violation is given by the area of the Jarlskog triangle. For any Jarlskog triangle, the area is the same. In the Wolfenstein parametrization \cite{Wol83}, the area of the Jarlskog triangle is of order $\lambda^6$. In Fig. \ref{fig:JTriangle}, we show the triangle with two long sides of order $\lambda$. Rotating the $O(\lambda^5)$ side (the red arrow), the CP phase $\delta$ and also the area change. From  Fig. \ref{fig:JTriangle}, we note that the area is maximum for  $\delta\simeq\frac{\pi}{2}$, and the maximality $\delta=\frac\pi{2}$ is a physical statement. The maximal CP violation can be modeled as recently shown in \cite{Kim11plb}.

%%%%%%%%%%%%%%%%%%%%%%%%%%%%%%%%%%%%%%%%%%%%%%%%%%%%%%%%%%%%%%%%%%%%%%%%%%%%
\acknowledgments{This work is supported in part by the National Research Foundation  (NRF) grant funded by the Korean Government (MEST) (No. 2005-0093841).}

\vskip 0.5cm
%%%%%%%%%%%%%%%%%%%%%%%%%%%%%%%%%%%%%%%%%%%%%%%%%%%%%%%%%%%%%%%%%%%%%%%%%%

\end{document}